\documentclass[12pt]{iopart}
\usepackage{graphicx}

\begin{document}

\title{Memory Effects and Transport Coefficients for Non-Newtonian Fluids}
\author{T. Kodama$^1$ and T. Koide$^2$}

\begin{abstract}
We discuss the roles of viscosity in relativistic fluid dynamics from the
point of view of memory effects. Depending on the type of
quantity to which the memory effect is applied, different terms appear in
higher order corrections. We show that when the memory effect applies on the
extensive quantities, the hydrodynamic equations of motion become
non-singular. We further discuss the question of memory effect in the
derivation of transport coefficients from a microscopic theory. We
generalize the application of the Green-Kubo-Nakano (GKN) to calculate
transport coefficients in the framework of projection operator formalism,
and derive the general formula when the fluid is non-Newtonian.
\end{abstract}

\address{$^1$ Instituto de F\'{\i}sica, Universidade Federal do Rio de Janeiro, C. P.
68528, 21945-970, Rio de Janeiro, Brazil}
\address{$^2$ Frankfurt Institute for Advanced Study - FIAS, Frankfurt, Germany}

% \maketitle

\section{Non-Newtonian Nature of a Dissipative Fluid in Relativistic Regime}

The effect of dissipation in relativistic fluids is one of current topics in
the physics of relativistic heavy-ion collisions \cite{Viscous}. Here, we
discuss this problem focusing on the memory effect on irreversible current.
It is well-known that in a simple covariant extension of the Navier-Stokes
theory there appears the problem of relativistic acausality and instability
associated with it. First let us illustrate using the example of diffusion
equation, the basic idea of memory effect as the solution for the problem of
acausality in relativistic hydrodynamics.

In the usual derivation of diffusion equation, we assume that the
irreversible current $J(t)$ is simply proportional to the corresponding
thermodynamic force $F(t)$, 
\begin{equation}
J(t)=DF(t),  \label{eqn:n-current}
\end{equation}%
where $D$ is a transport coefficient. The fluid whose irreversible current
posseses this property is referred to as a Newtonian fluid, and the
evolution is described by, for example, the Navier-Stokes equation. However,
to be exact, there should exist some time retardation effect in generating
the current in the medium, when a thermodynamic force is applied (remember
the linear response theory). The expression (\ref{eqn:n-current}) is
justified only when there is clear separation of microscopic and macroscopic
scales and the retardation effect is negligible. This assumption is usually
satisfied in fluid around us, where the average velocity of molecules is
larger than the velocity of the sound, but the speed of a fluid diffusion is
much slower.

However, depending on the fluid, there exist many counter examples of this
Newtonian behavior. There, the clear separation of microscopic and
macroscopic timescales disappear due to the correlations, and we should
consider the retardation effect in the definition of relativistic
irreversible currents, 
\begin{equation}
J(t)=\int_{-\infty }^{t}G(t-s)F(s),  \label{eqn:n-current2}
\end{equation}%
where $G(t)$ is a memory function which represents the retardation effect.

Another situation where the Newtonian nature is not obvious is the motion of
viscous fluids in relativistic regime. There, the fluid can be accelerated
up to the speed of light and the proper time is slowed down due to the
Lorentz retardation. Therefore, the effect of retardation, even small in
non-relativistic regime, becomes crucial in the dynamics of a viscous fluid.
Therefore, to include the retardation effects in the formulation of viscous
hydrodynamics is a natural step to extend the fluid dynamics in the
relativistic regime and, as a matter of fact, this extension is crucially
related to the problem of acausality as seen below.

Let us consider a diffusion process \cite{koide1}. Then the thermodynamic
force is given by the spatial derivative of a conserved density $n$ and the
linear relation (\ref{eqn:n-current}) is called Fick's law. By substituting
Fick's law into the equation of continuity, we obtain the diffusion
equation. As is well known, the diffusion equation has a problem of infinite
propagation speed. On the other hand, when we use Eq. (\ref{eqn:n-current2})
with a simple exponential memory function, 
\begin{equation}
G(t)=\frac{D}{\tau _{R}}e^{-t/\tau _{R}},
\end{equation}%
where $\tau _{R}$ is the relaxation time and $D$ is the diffusion constant,
we obtain a telegraph equation and then the maximum propagation speed is
given by $\sqrt{D/\tau _{R}}$. It is also worth for mentioning that what we
can obtain from microscopic dynamics is not the diffusion equation but the
telegraphic equation \cite{koide1}.

The same discussion is applicable to relativistic dissipative hydrodynamics 
\cite{dkkm1}. In the relativistic Navier-Stokes (Landau-Lifshitz) equation,
the viscous flows are defined by assuming Eq. (\ref{eqn:n-current}). For
example, in the bulk viscosity, the corresponding thermodynamic force is $%
\partial _{\mu }u^{\mu }$, where $u^{\mu }$ is the fluid velocity. Thus the
bulk viscosity is given by 
\begin{equation}
\Pi =-\eta \partial _{\mu }u^{\mu }.  \label{eqn:bulk_ns}
\end{equation}%
On the other hand, in the hydrodynamics consistent with causality, it is
given by 
\begin{equation}
\pi (\tau )=\int_{-\infty }^{\tau }G(\tau -s)\partial _{\mu }u^{\mu }(s),
\label{eqn:bulk_cdh}
\end{equation}%
where $\tau $ is a proper time. As is shown in Ref. \cite{dkkm2,dkkm3}, the
propagation speed of signal exceeds the speed of light in the case of Eq. (%
\ref{eqn:bulk_ns}), and this acausality makes the hydrodynamic evolution
unstable. On the other hand, hydrodynamic motion with a covariant
formulation of the retarded irreversible current (\ref{eqn:bulk_cdh})
becomes causal and stable. Due to the acausality and associated instability
problem, it is impossible to apply the Landau-Lifshitz theory in a physical
system without using artificial tricks (such as to use the value of $%
\partial _{\mu }u^{\mu }$ of the former step in the time-integration
algorithm).

\section{Memory Effects on Extensive Measures}

Another important mechanism is the finiteness of fluid cells to be applied
thermodynamic relations \cite{dkkm4}. In the derivation of hydrodynamics,
it is assumed that the local equilibrium is achieved in each fluid cells
which has finite spatial extension (to be rigorous, it should be infinite
compared with the microscopic scale). Then, the second law of thermodynamics
and the memory effect should be applied on extensive measures associated
with this finite volume. To introduce an extensive measure for the density
of an additive quantity, let us consider a fluid cell of proper volume $V.$
Due to the fluid flow, this volume changes in time and its time rate of
change is given by%
\[
\frac{1}{V}\frac{dV}{dt}=\nabla \cdot \vec{v},
\]%
where $\vec{v}$ is the fluid velocity field. This equation can be written in
a covariant form as%
\begin{equation}
\partial _{\mu }(\sigma (\mathbf{r},t)u^{\mu }(\mathbf{r},t))=0,
\end{equation}%
where $\sigma (\mathbf{r},t)$ is the inverse of the volume of the fluid cell
at $\mathbf{r}$, and $u^{\mu }$ is the fluid velocity. Then the irreversible
current is given by 
\begin{equation}
\frac{J(t)}{\sigma (t)}=\int_{t_{0}}^{t}dsG(t-s)\frac{F(s)}{\sigma (s)}
\end{equation}%
This is equivalent to the solution of the following differential equation, 
\begin{equation}
\tau _{R}\frac{d}{dt}J(t)+J(t)=DF(t)-\tau _{R}J(t)\partial _{\mu }u^{\mu
}(t).  \label{eqn:bulk_fve}
\end{equation}%
Noted that this result does not depend on the choice of the volume of the
fluid cell $\sigma (t)$. The above equation differs from that of the
so-called Maxwell-Cattaneo equation (or in fluid mechanics, the truncated Israel-Stewart theory), 
\begin{equation}
\tau _{R}\frac{d}{dt}J(t)+J(t)=DF(t)  \label{Tranc_IS}
\end{equation}%
which can also be obtained from the memory effect applied on the current
density itself as%
\[
J(t)=\int_{t_{0}}^{t}dsG(t-s)F(s).
\]

The last non-linear term in (\ref{eqn:bulk_fve}) is important not only for
the physical concept of the irreversible current, but also for the stability
of relativistic fluids. Actually there are two different origins for
instabilities in relativistic dissipative hydrodynamics. One is the
instability induced by acausality which we addressed in the previous section
(Here we would like to call attention that, it is sometimes considered that
the problems of acausality and instability are different problems but as
shown in Ref. \cite{dkkm3}, the instability of relativistic fluids is
induced by acausality itself), which can be removed even by the truncated
Israel-Stewart theory (\ref{Tranc_IS}). The other instability appears in
ultra-relativistic phenomena even the theory is causal. When the term $%
\partial _{\mu }u^{\mu }$ becomes very large, the bulk viscosity $\Pi $ can
take a very large negative value. In such extreme situation, we can
analytically show that the mass matrix associated with the simple causal
dissipative hydrodynamics Eq.(\ref{Tranc_IS}) becomes negative, hence  it is
unstable \cite{dkkm4}.

On the other hand, such a instability never appears when we consider the
non-linear term in Eq.(\ref{eqn:bulk_fve}). This is because, the presence of
the non-linear term prevents the bulk viscosity to become negatively very
large. We found that there exist the minimum value of $\Pi $, given by \cite%
{dkkm4} 
\begin{equation}
\Pi \geq -\frac{\zeta }{\tau _{R}},  \label{Pi_min}
\end{equation}%
where $\zeta $ is the bulk viscosity coefficient and $\tau _{R}$ is the
corresponding relaxation time. Due to this condition, the mass matrix of the
hydrodynamic equation is proved to be positive definite. Thus the causal
dissipative hydrodynamics with finite size effect is stable even for
ultra-relativistic phenomena. 

Another important fact is that, the so-called full Israel-Stewart theory can
also be expressed in terms of a memory function. However, in this case a
very peculiar form of thermodynamic quantity should be taken to apply the
memory effect and the physical interpretation from the extended
thermodynamic point of view is difficult \cite{dkkm4}.

\begin{figure}[ptb]
\begin{minipage}{.45\linewidth}
\includegraphics[scale=0.3]{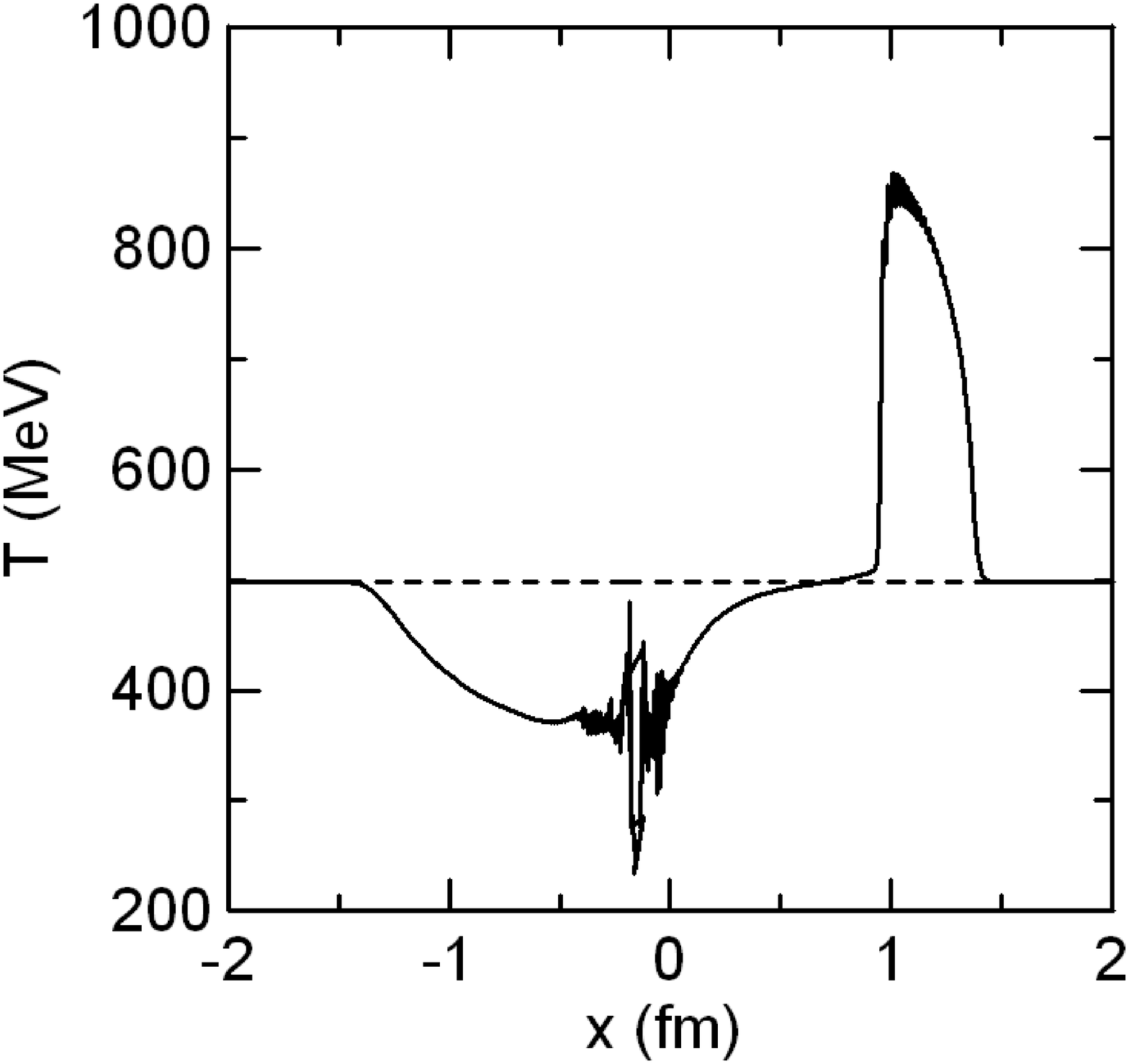}
%\caption{The temperature in the shock formation calculated in LCDH with $a=1$ at $t=0.75$ fm,
%starting from the homogeneous initial condition (dotted line).}
%\label{temp}
\end{minipage}
\begin{minipage}{.45\linewidth}
\includegraphics[scale=0.3]{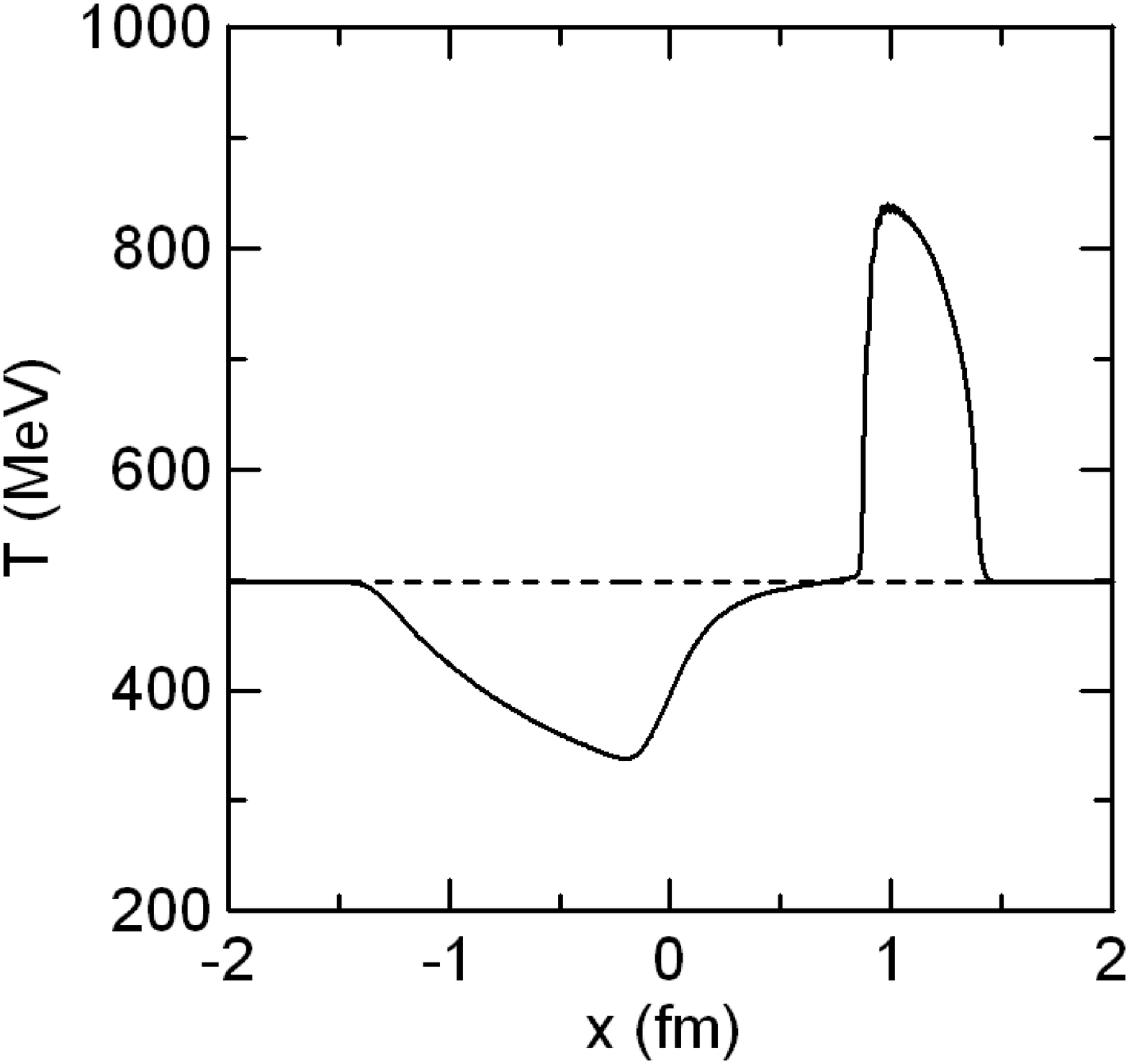}
%\caption{The temperature in the shock formation calculated without the finite size effect with $a=1$ and $b=6$ at $t=0.75$ fm,
%starting from the homogeneous initial condition (dotted line).}
%\label{temp2}
\end{minipage}
\caption{The temperature in the shock formation calculated without the
finite size effect (left) and with the finite size effect (right), starting
from the homogeneous initial condition (dotted line) \protect\cite{dkkm4}.}
\label{temp}
\end{figure}

In Fig. \ref{temp}, we show the shock formation with an initial velocity $%
\gamma = 5$. As is shown in the left panel, in this ultra-relativistic
initial condition, the causal dissipative hydrodynamics without the finite
size effect becomes unstable. On the other hand, if the finite size effect
is taken into account, the instability disappears.

A very different scenario where relativistic fluid dynamics can be applied
is found in astrophysics. The speed of flow becomes $\gamma \sim 100$ in the
Gamma-raybursts. Thus, the construction of the dissipative hydrodynamics
applicable to such an extreme situation is important for some astrophysical
processes, too.

\section{GKN formula : Applicable to non-Newtonian Fluids ?}

As we have shown, it is natural to consider that the behavior of a fluid in
relativistic regime is non-Newtonian type. This means that we cannot use
various techniques which are developed to analyze the Newtonian fluids. The
Green-Kubo-Nakano (GKN) formula to derive the transport coefficients is one
of them. The GKN formula has been applied to calculate the viscosity
coefficients and the heat conductivity of the quark-gluon plasma. However,
it should be noted that the assumption of Newtonian fluids is implicitly
used in the derivation of the expressions. Thus those values of transport
coefficient obtained from the usual GKN formula might not be applicable to
be used in the relativistic regime. We have to consider how the
non-Newtonian nature affects the values of transport coefficients.

To see this, we consider the system whose Hamiltonian is given by $H$. By
applying an external force, the total Hamiltonian is changed from $H$ to $%
H+H_{ex}(t)$, with 
\begin{equation}
H_{ex}(t)=-AF(t),
\end{equation}%
where $A$ is an operator and $F(t)$ is the c-number external force.

We consider the current $J$ induced by the external force. In the framework
of the linear response theory, we can write 
\begin{equation}
\langle J\rangle =\int_{-\infty }^{t}ds\Psi (t-s)F(s),  \label{eqn:GKN_ori}
\end{equation}%
where the response function is given by 
\begin{equation}
\Psi (t)=\int_{0}^{\beta }d\lambda \langle \dot{A}(-i\lambda )J(t)\rangle
_{eq},
\end{equation}%
where $\beta =1/T$ and $T$ is the temperature. This is the exact result in
the sense of the linear approximation and one of the important expressions
of the GKN formula. However, when we define transport coefficients of
hydrodynamics for the non Newtonian fluid, we need to take care to use this
expression.

In applying the GKN formula for hydrodynamic transport coefficients, we
assume the relation between currents and the external force such as  $%
J(t)=D_{GKN}F(t)$ which define the transport coefficient $D_{GKN}$. One can
immediately see that this is nothing but the assumption of Newtonian fluids.
Commonly the formula to calculate this $D_{GKN}$ in this way is usually
called the \textquotedblleft GKN formula\textquotedblright\ of the shear
viscosity, the bulk viscosity, heat conduction and so on. To derive the
expression for $D_{GKN}$, we have to ignore the memory effect
(time-convolution integral) in Eq. (\ref{eqn:GKN_ori}), 
\begin{equation}
\langle J\rangle \approx \int_{0}^{\infty }ds\Psi (s)F(t).
\end{equation}%
Then the GKN formula is reduced to 
\begin{equation}
D_{GKN}=\int_{0}^{\infty }ds\Psi (s).
\end{equation}%
In the case of shear viscosity, this is given by the time correlation
function of energy-momentum tensor.

In principle, we can derive the formula for non-Newtonian transport
coefficients by assuming Eq.(\ref{eqn:bulk_fve}) instead of Eq.(\ref%
{eqn:n-current}). From the exact result Eq.(\ref{eqn:GKN_ori}), we can
derive the following equation, 
\begin{equation}
\partial _{t}J(t)=\Psi (0)F(t)+\int_{0}^{\infty }ds\partial _{s}\Psi (s)F(t).
\end{equation}%
In the second term, we ignore the time-convolution integral. In the first
order approximation in the relaxation time, we may use the usual GKN formula
to re-express the first term. Then we finally obtain 
\begin{equation}
\partial _{t}J(t)=\frac{\Psi (0)}{D_{GKN}}J(t)+\int_{0}^{\infty }ds\partial
_{s}\Psi (s)F(t).
\end{equation}%
By comparing this equation with Eq. (\ref{eqn:bulk_fve}) ignoring the
non-linear term, we can derive the expression of $D$ and $\tau _{R}.$ The
results obtained in this way differ from those of the direct application of
GKN formulas to Newtonian fluid. However, in the limit of Markovian
approximation of the correlation function, the two results coincide.

Strictly speaking, the shear viscosity is induced not by the external force
but by the difference of the boundary conditions. Actually, the GKN formula
of the shear viscosity is derived by using the nonequilibrium statistical
operator method proposed by Zubarev. Thus the discussion we developed here
is not directly applicable to the problems discussed in this paper. More
detailed discussion and the exact expression of the non-Newtonian transport
coefficients are given in \cite{koide2,koide3}.

We have investigated the viscous fluid dynamics emphasizing the memory
effect. When a fluid posses a non-Newtonian behavior, the use of GKN formula
should be cautious.

We thank fruitful discussion with G.Denicol and Ph. Mota on the extensivity of the memory effects. 
This work has been supported by PRONEX, FAPERJ and CNPq.

\end{document}